\documentclass[10pt]{npqcd}

\begin{document}
\renewcommand\nextpg{\pageref{pgs1}}
\renewcommand\titleA{
 Towards thermodynamics of the quark quasi-particles
}
\renewcommand\authorA{
 S. V. Molodtsov${}^{1,2}$${}^{\mathrm{a}}$,
 G. M. Zinovjev${}^3$${}^{\mathrm{b}}$
}
\renewcommand\email{
 e-mail:\space \eml{a}{molodtsov@itep.ru},
\eml{b}{Gennady.Zinovjev@cern.ch}\\[1mm]
}
\renewcommand\titleH{
 Towards thermodynamics of the quark quasi-particles
}
\renewcommand\authorH{
 Molodtsov S.V., Zinovjev G.M.
}
\renewcommand\titleC{\titleA}
\renewcommand\authorC{\authorH}
\renewcommand\institution{
${}^1$Joint Institute for Nuclear Research, Dubna,
Moscow region, RUSSIA\\
${}^2$Institute of Theoretical and Experimental Physics, Moscow, RUSSIA\\
${}^3$Bogolyubov Institute for Theoretical Physics,
National Academy of Sciences of Ukraine, Kiev, UKRAINE
}
\renewcommand\abstractE{
Some features of hot and dense gas of quarks which are considered as the
quasi-particles of the model Hamiltonian with four-fermion interaction
are studied. Being adapted to the Nambu-Jona-Lasinio model this approach
allows us to accommodate a phase transition similar to the nuclear liquid-gas
one at the proper scale and to argue an existence of the mixed phase of vacuum and
normal baryonic matter as a plausible scenario of chiral
symmetry (partial) restoration. Analyzing the transition layer
between two phases we estimate the surface tension coefficient and
discuss the possibility of quark droplet formation.
}
\begin{article}

Understanding in full and describing dependably the critical phenomena (chiral
and deconfinement phase transitions) in QCD is still elusive because of a
necessity to have the corresponding efficient non-perturbative methods for
strongly coupled regime. For the time being such studies are pursued by invoking
diverse effective models. The Nambu-Jona-Lasinio(NJL)-type models are certainly
playing the most advanced role in this analysis \cite{molodtsov:Bub}. This
approach deals with the multi-fermion interactions in lieu of a gluon field QCD
dynamics and does not incorporate the property of confinement. At the same time
it is quite successful in realizing the spontaneous breakdown of chiral symmetry
and its restoration at nonzero temperatures or quark densities.

These and some related items are discussed in this paper inspired by well known
and fruitful idea about the specific role of surface degrees of freedom in the
finite fermi-liquid systems and in considerable extent by our previous works
\cite{molodtsov:MZ} and \cite{molodtsov:ff} in which the quarks were treated as
the quasi-particles of the model Hamiltonian and the problem of filling up the
Fermi sphere was studied in detail. Under such a treatment an unexpected
singularity (discontinuity) of the mean energy functional as a function of the
current quark mass was found. In the particular case of the NJL model the
existence of new solution branches of the equation for dynamical quark mass as a
function of chemical potential have been demonstrated and the appearance of
state filled up with quarks which is almost degenerate with the vacuum state both in
the quasi-particle chemical potential and in the ensemble pressure has been
discovered.

Here we are going to study the quark ensemble features at finite temperature and
fixed baryonic chemical potential and to analyse the first order phase
transition which takes place in such a system of free quasi-particles. Analysis is
performed within the framework of two approaches which are supplementary, in a sense,
albeit giving the identical results. One of those approaches, based on the Bogolyubov
transformations, is especially informative to study the process of filling the
Fermi sphere up because at this point the density of quark ensemble develops a
continuous dependence on the Fermi momentum. It allows us to reveal an additional structure
in the solution of gap equation for dynamical quark mass just in the proper
interval of parameters characteristic for phase transition and to trace its evolution. The
result is that a quark ensemble might be found in two aggregate states, gas and liquid,
and the chiral condensate is partially restored in a liquid phase. In order to make
these conclusions easily perceptible we deal with the simplest version of the NJL
model (with one flavor and one of the standard parameter sets) and, actually, do not
aim to adjust the result obtained with well-known nuclear liquid-gas phase transition.
Besides, it seems our approach might be treated as a sort of microscopic ground
of the conventional bag model and those states filled up with quarks are
conceivable as a natural 'construction material' for baryons.

Now as an input to start with we remind the key elements of approach developed.
The corresponding Hamiltonian includes the interaction term taken in the form of a
product of two coloured currents located in the spatial points ${\bf x}$ and
${\bf y}$ which are connected by a form-factor and its density reads as
\begin{equation}
\label{molodtsov:1}
{\cal H}=-\bar q(i{\bf \gamma}{\bf \nabla}+im)q-\bar q t^a\gamma_\mu q
\int d{\bf y}\bar q' t^b\gamma_\nu q' \langle A^{a}_\mu A'^{b}_\nu\rangle,
\end{equation}
where $q=q({\bf x})$, $\bar q=\bar q({\bf x})$, $q'=q({\bf y})$,
$\bar q'=\bar q({\bf y})$ are the quark and anti-quark operators,
\begin{equation}
\label{molodtsov:2}
q_{\alpha i}({\bf x})=\int\frac{d {\bf p}}{(2\pi)^3} \frac{1}{(2|p_4|)^{1/2}}
\left[a({\bf p},s,c)u_{\alpha i}({\bf p},s,c) e^{i{\bf p}{\bf x}}
+b^+({\bf p},s,c)v_{\alpha i}({\bf p},s,c) e^{-i{\bf p}{\bf x}}\right],
\nonumber
\end{equation}
$p_4^2=-{\bf p}^2-m^2$, $i$--is the colour index, $\alpha$ is the spinor index
in the coordinate space, $a^+$, $a$ and $b^+$, $b$ are the creation and annihilation
operators of quarks and anti-quarks, $a|0\rangle=0$, $b|0\rangle=0$, $|0\rangle$ is the
vacuum state of free Hamiltonian and $m$ is a current quark mass. The summation over
indices $s$ and $c$ is meant everywhere, the index $s$ describes two spin polarizations of
quark and the index $c$ plays the similar role for a colour. $t^a=\lambda^a/2$ are the
generators of $SU(N_c)$ colour gauge group, the Hamiltonian density is considered in the
Euclidean space and $\gamma_\mu$ denote the Hermitian Dirac matrices, $\mu,\nu=1,2,3,4$.
$\langle A^{a}_\mu A'^{b}_\nu\rangle$ stands for the ~form-factor of the
following form $\langle A^{a}_\mu A'^{b}_\nu\rangle=\delta^{ab}
 \frac{2~\widetilde G}{N_c^2-1}\left[I({\bf x}-{\bf y})
\delta_{\mu\nu}-J_{\mu\nu}({\bf x}-{\bf y})\right]$,
where the second term is spanned by the relative distance vector and the gluon
field primed denotes that in the spatial point ${\bf y}$. The effective Hamiltonian
density (\ref{molodtsov:1}) results from averaging the ensemble of quarks influenced by
intensive stochastic gluon field $A^a_\mu$, see Ref. \cite{molodtsov:MZ}. For the sake of
simplicity in what follows we neglect the contribution of the second term of
Eq.(\ref{molodtsov:1}).
The ground state of the system is searched as the Bogolyubov trial function
composed of the quark-anti-quark pairs with opposite momenta and with vacuum quantum numbers, i.e.
\begin{equation}
\label{molodtsov:4}
|\sigma\rangle={\cal{T}}~|0\rangle~,~~~
{\cal{T}}=\Pi_{ p,s}\exp\{\varphi~[a^+({\bf p},s)b^+(-{\bf p},s)+
a({\bf p},s)b(-{\bf p},s)]\}.\nonumber
\end{equation}
In this formula and below, in order to simplify the notations we refer to one
compound
index only which means both the spin and colour polarizations. The parameter
$\varphi({\bf p})$ which describes the pairing strength is determined by the
minimum of
mean energy $E=\langle\sigma|H|\sigma\rangle$. By introducing the 'dressing
transformation'
we define the creation and annihilation operators of quasi-particles as
$A={\cal{T}}a{\cal{T}}^{-1}$, $B^+={\cal{T}}b^+{\cal{T}}^{-1}$ and for fermions
${\cal{T}}^{-1}={\cal{T}}^\dagger$. Then the quark field operators are presented
as
\begin{eqnarray}
\label{molodtsov:6}
&&q({\bf x})=\int\frac{d {\bf p}}{(2\pi)^3} \frac{1}{(2|p_4|)^{1/2}}~
\left[~A({\bf p},s)~U({\bf p},s)~e^{i{\bf p}{\bf x}}+
B^+({\bf p},s)~V({\bf p},s)~ e^{-i{\bf p}{\bf x}}\right]~,\nonumber\\
&&\bar q({\bf x})=\int\frac{d {\bf p}}{(2\pi)^3} \frac{1}{(2|p_4|)^{1/2}}~
\left[~A^+({\bf p},s)~\overline{U}({\bf p},s)~e^{-i{\bf p}{\bf x}}+
B({\bf p},s)~\overline{V}({\bf p},s)~ e^{i{\bf p}{\bf x}}\right]~,
\nonumber
\end{eqnarray}
and the transformed spinors $U$ and $V$ are given by the following forms
$U({\bf p},s)=\cos(\varphi)u({\bf p},s)-
\sin(\varphi)v(-{\bf p},s)$,
$V({\bf p},s)=\sin(\varphi)u(-{\bf p},s)+
\cos(\varphi)v({\bf p},s)$
where $\overline{U}({\bf p},s)=U^+({\bf p},s)\gamma_4$,
$\overline{V}({\bf p},s)=V^+({\bf p},s)\gamma_4$ are the Dirac conjugated
spinors.

In Ref. \cite{molodtsov:ff} the process of filling in the Fermi sphere with the
quasi-particles of quarks was studied by constructing the state of the Sletter
determinant type
$|N\rangle=\prod_{|{\mbox{\scriptsize{\bf P}}}|<P_F;S}A^+
({\bf P};S)|\sigma\rangle$
which possesses the minimal mean energy over the state $|N\rangle$. The
polarization indices run through all permissible values here and the quark
momenta are bounded by the limiting Fermi momentum $P_F$. The momenta and
polarizations of states forming the quasi-particle gas are marked by the capital
letters similar to above formula and the small letters are used in all other
cases.

As it is known the ensemble state at finite temperature $T$ is described by the
equilibrium statistical operator $\rho$. Here we use the Bogolyubov-Hartree-Fock
approximation in which the corresponding statistical operator is presented by
the following form
\begin{equation}
\label{molodtsov:dm}
\rho=\frac{e^{-\beta ~\hat H_{{\mbox{\scriptsize{app}}}}}}{Z_0}~,
~~Z_0=\mbox{Tr}~\{e^{-\beta ~\hat H_{{\mbox{\scriptsize{app}}}}}\}~~,
\end{equation}
where an approximating effective Hamiltonian $H_{{\mbox{\scriptsize{app}}}}$ is
quadratic in the creation and annihilation operators of quark and anti-quark
quasi-particles $A^+$, $A$, $B^+$, $B$ and is defined in the corresponding Fock
space with the vacuum state $|\sigma\rangle$ and $\beta=T^{-1}$. There is no
need to know the exact form of this operator henceforth because all the quantities of
our interest in the Bogolyubov-Hartree-Fock approximation are expressed by the
corresponding averages (a density matrix)
$n(P)=\mbox{Tr} \{\rho~ A^+({\bf P};S) A({\bf P};S)\}$,
$\bar n(Q)=\mbox{Tr} \{\rho~ B^+({\bf Q};T) B({\bf Q};T)\}$,
which are found by solving the following variational problem. The statistical
operator $\rho$ is determined in such a form in order to have at the fixed mean charge
\begin{equation}
\label{molodtsov:ntot}
\bar Q_4=\mbox{Tr} \{\rho ~Q_4\}=
V ~2 N_c \int  \frac{d {\bf p}}{(2\pi)^3}~ [n(p)-\bar n(p)]~,
\end{equation}
where
$Q_4=\int\frac{d{\bf p}}{(2\pi)^3}\frac{-
ip_4}{|p_4|}\left[A^+(p)A(p)+B(p)B^+(p)\right]$
(for the diagonal component of our interest here,
$Q_4=-\int d{\bf x} \bar q i \gamma_4 q$) and fixed mean entropy
\begin{eqnarray}
\label{molodtsov:stot}
&&\bar S=-\mbox{Tr} \{\rho \ln \rho\}=\nonumber\\[-.2cm]
\\ [-.25cm]
&&=-V 2 N_c \int\!\!\!  \frac{d {\bf p}}{(2\pi)^3}
\left[n(p)\ln n(p)+(1-n(p))\ln (1-n(p))
+\bar n(p)\ln \bar n(p)+(1-\bar n(p))\ln (1-\bar n(p))\right],\nonumber
\end{eqnarray}
($S=-\ln \rho$) the minimal value of mean energy of quark ensemble
$E=\mbox{Tr} \{\rho H\}$. The definition of mean charge is given here up to the
unessential (infinite) constant coming from permuting the operators $B B^+$ in
the charge operator $Q_4$. It is reasonable to remind that the mean charge should be
treated in some statistical sense because it characterizes quark ensemble
density and has no colour indices.

Calculating the corresponding matrix elements leads to the following result for
mean energy density per one quark degree of freedom (the details can be found
in \cite{molodtsov:wear}) $w={\cal E}/{2 N_c}$, ${\cal E}=E/V$ where $E$ is a
total ensemble energy
\begin{eqnarray}
\label{molodtsov:15}
w&=&\int\frac{d {\bf p}}{(2\pi)^3} |p_4| +
\int\frac{d {\bf p}}{(2\pi)^3}|p_4|\!\cos\theta[n(p)+\bar n(p)-1]-
\nonumber
\\[-.2cm]\\ [-.25cm]
&-&G\int \frac{d {\bf p}}{(2\pi)^3}\sin \left(\theta-\theta_m\right)
[n(p)+\bar n(p)-1]\int \frac{d {\bf q}}{(2\pi)^3}\sin\left(\theta'-
\theta'_m\right)
[n(q)+\bar n(q)-1]~I~.\nonumber
\end{eqnarray}
(up to the constant unessential for our consideration here). Here the following
denotes are used $p=|{\bf p}|$, $q=|{\bf q}|$, $\theta=2 \varphi$, $\theta'=\theta(q)$,
$I= I({\bf p}+{\bf q})$ and the angle $\theta_m(p)$ is determined by
$\sin \theta_m=m/|p_4|$. It was quite practical to single
 out the colour factor in the four-fermion coupling constant as $G=2 \widetilde G/N_c$.
It is of important to notice that the existence of such
an angle stipulates the discontinuity of mean energy functional mentioned above and
found out in \cite{molodtsov:MZ}

We are interested in minimizing the following functional
$\Omega=E-\mu~\bar Q_4 -T~\bar S$ where $\mu$ and $T$ are the Lagrange factors
for the chemical potential and temperature respectively. The approximating Hamiltonian
$\hat H_{{\mbox{\scriptsize{app}}}}$ is constructed simply by using the
information on $E-\mu~\bar Q_4$ of presented functional (see, also below).
For the specific contribution per one quark degree of freedom $f=F/2N_c$,
$F=\Omega/V$ we receive
\begin{eqnarray}
\label{molodtsov:17}
&&f=\int \frac{d{\bf p}}{(2\pi)^3}
\left[|p_4|\cos\theta(n+\bar n -1)-\mu(n-\bar n)\right]
+\int \frac{d{\bf p}}{(2\pi)^3}|p_4|-G\int \frac{d{\bf p}}{(2\pi)^3}
\sin \left(\theta-\theta_m\right)(n+\bar n-1)\times\nonumber\\[-.2cm]
\\ [-.25cm]
&&\times\int\frac{d{\bf q}}{(2\pi)^3}\sin\left(\theta'-\theta'_m\right)
(n'+\bar n'-1)I+T\int \frac{d{\bf p}}{(2\pi)^3}
\left[n\ln n+(1-n)\ln(1-n)+\bar n\ln \bar n+(1-\bar n)\ln(1-\bar
n)\right].\nonumber
\end{eqnarray}
Here the primed variables correspond to the momentum $q$. The optimal values of
parameters are determined by solving the following system of equations
($df/d\theta=0$, $df/d n=0$, $df/d \bar n=0$)
\begin{eqnarray}
\label{molodtsov:18}
&&|p_4|~\sin\theta-M\cos \left(\theta-\theta_m\right)=0~,\nonumber\\[-.2cm]
\\ [-.25cm]
&&|p_4|~\cos\theta-\mu+M~\sin \left(\theta-\theta_m\right)-
T~\ln \left(n^{-1}-1\right)=0~,\nonumber\\
&&|p_4|~\cos\theta+\mu+M~\sin \left(\theta-\theta_m\right)-
T~\ln \left(\bar n^{-1}-1\right)=0~\nonumber
\end{eqnarray}
where we denoted the induced quark mass as
\begin{equation}
\label{molodtsov:19}
\hspace{-0.3cm}
M({\bf p})=\!\!2G\!\!\int\!\!\! \frac{d{\bf q}}{(2\pi)^3}
(1-n'-\bar n')\sin \left(\theta'-\theta'_m\right)I({\bf p}+{\bf q}).
\end{equation}

Turning to the presentation of obtained results in the form customary for mean
field approximation we introduce a dynamical quark mass $M_q$ parameterized as
$\sin \left(\theta-\theta_m\right)=\frac{M_q}{|P_4|}$,
$|P_4|=({\bf p}^2+M_q^{2}({\bf p}))^{1/2}$
and ascertain the interrelation between induced and dynamical quark masses. From
the first equation of system (\ref{molodtsov:18}) we fix the pairing angle as
$\sin \theta=p M/(|p_4||P_4|)$ and making use of the identity
$$(|p_4|^2-M ~m)^2+M^2 p^2=[p^2+(M-m)^2]~|p_4|^2~$$
find out that
$\cos \theta=\pm\frac{|p_4|^2-m~M}{|p_4||P_4|}$. For clarity we choose the upper
sign 'plus'. Then, as an analysis of the NJL model teaches, the branch of
equation solution for negative dynamical quark mass is the most stable one. Let us
remember here that we are dealing with the Euclidean metrics (though it is not a
principal point) and a quark mass appears in the corresponding expressions as an imaginary
quantity. Now substituting the calculated expressions for the pairing angle into
the trigonometrical factor expression
$\sin \left(\theta-\theta_m\right)=
\sin \theta\frac{p}{|p_4|}-\cos \theta\frac{m}{|p_4|}$ and performing some
algebraic transformations of both parts of equation we define $M_q({\bf p})=M({\bf p})-m$.
And then the equation for dynamical quark mass (\ref{molodtsov:19}) is getting
the form characteristic for the mean field approximation
$$M=2G\int \frac{d{\bf q}}{(2\pi)^3}
(1-n'-\bar n')~\frac{M'_q}{|P'_4|}~I({\bf p}+{\bf q})~.$$
\begin{figure}
\begin{minipage}{.45\textwidth}
\centering
\includegraphics[bb= 0 0 600 600, width=.9\textwidth]{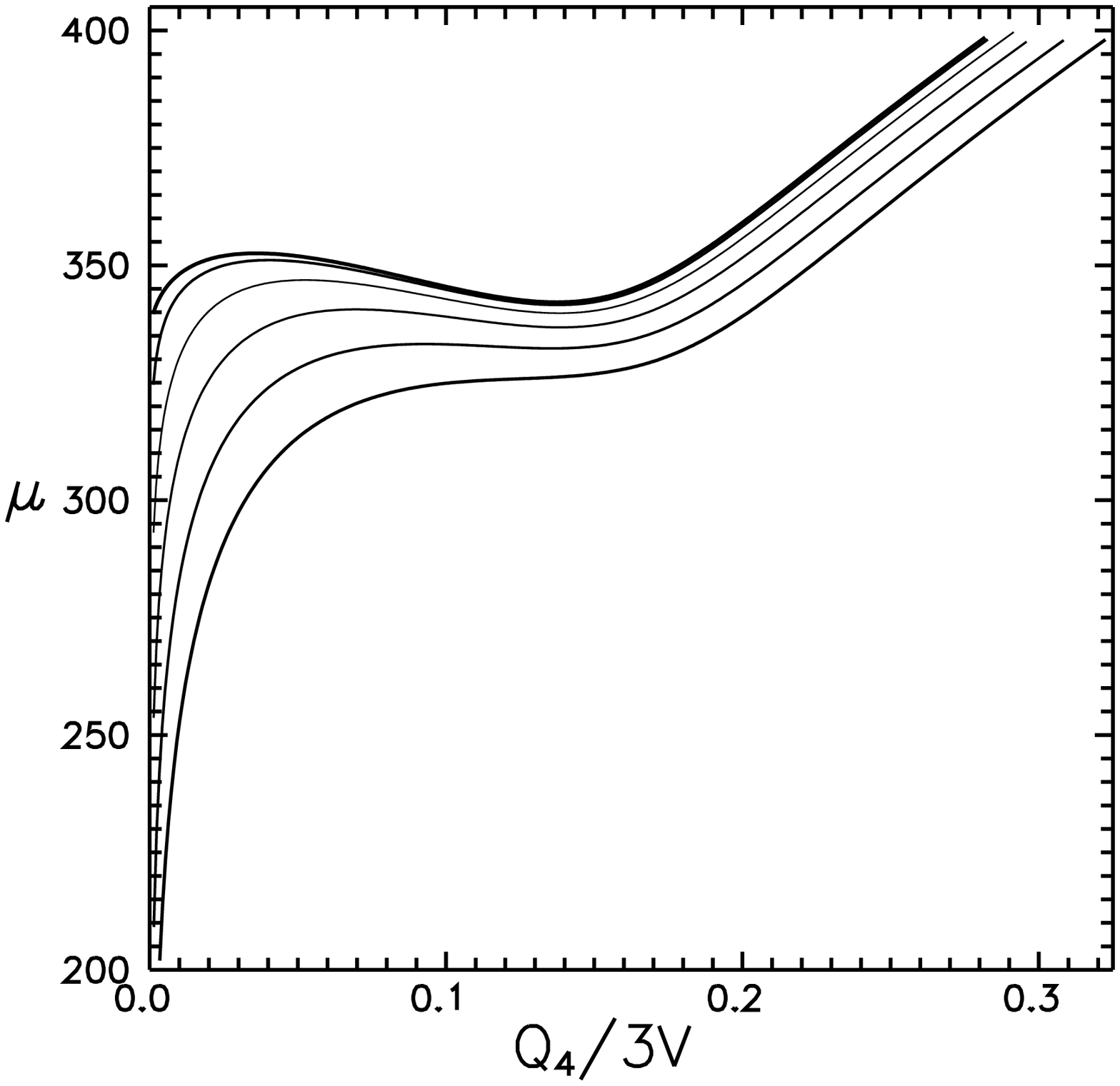}
\caption{The chemical potential $\mu$ (MeV) as a function of charge
density ${\cal Q}_4=Q_4/(3V)$ (in the units of charge/fm$^3$). The factor 3
relates the densities of quark and baryon matter. The top curve corresponds
to the zero temperature. The curves following down correspond to the temperature
values $T=10$ MeV, ... , $T=50$ MeV with spacing $T=10$ MeV.}
\label{molodtsov:f2}
\end{minipage}
 \rule{.05\textwidth}{0pt}
\begin{minipage}{.45\textwidth}
\centering
\includegraphics[bb= 0 0 600 600 ,width=.9\textwidth]{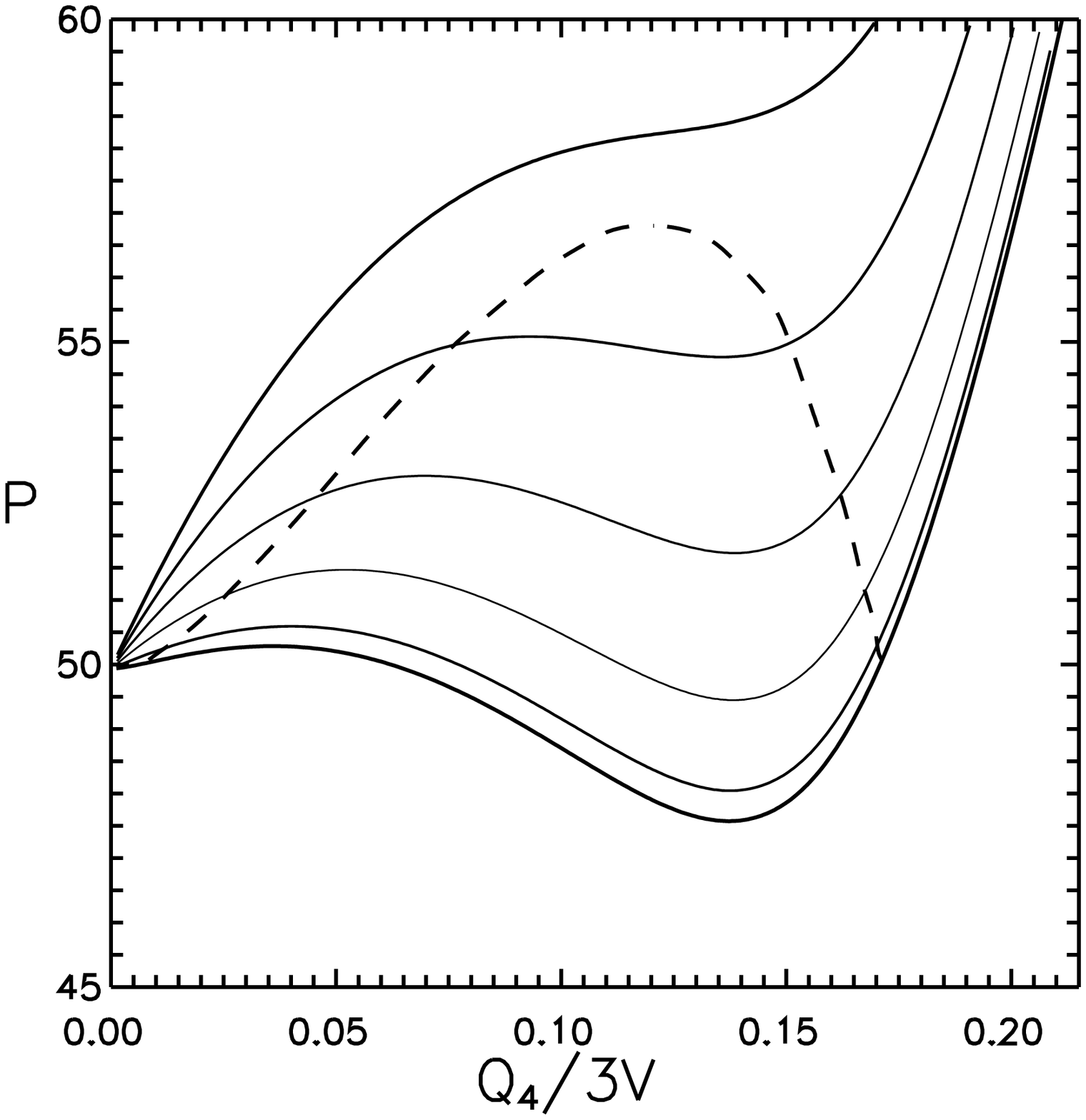}
\caption{The ensemble pressure $P$ (MeV/fm$^3$) as a function of charge density
${\cal Q}_4$ at temperatures $T=0$ MeV, ... , $T=50$ MeV with spacing $T=10$ MeV.
The lowest curve corresponds to zero temperature. The dashed curve shows the
boundary of phase transition liquid--gas, see the text.}
\label{molodtsov:f2a}
\end{minipage}
\end{figure}

The second and third equations of system (\ref{molodtsov:18}) allow us to find
for the equilibrium densities of quarks and anti-quarks as
$n=[e^{\beta~(|P_4|-\mu)}+1]^{-1}$,
$\bar n=[e^{\beta~(|P_4|+\mu)}+1]^{-1}$
and, hence, the thermodynamic properties of our system as well and, in
particular, the pressure of quark ensemble $P=-d E/d V$. By definition we should calculate
this derivative at constant mean entropy $d\bar S/dV=0$. This condition allows us,
for example, to calculate the derivative $d\mu/dV$. However, this way is not
reliable because then the mean charge $\bar Q_4$ might change, and it is more practical
to introduce two independent chemical potentials --- for quarks $\mu$ and for
anti-quarks $\bar\mu$ (following formula for $\bar n$ with an opposite sign). In
fact, it is the only possibility to obey both conditions simultaneously. It
leads to the following definitions of corresponding densities
$n=[e^{\beta~(|P_4|-\mu)}+1]^{-1}$,
$\bar n=[e^{\beta~(|P_4|+\bar\mu)}+1]^{-1}$.
On this way of description we are able even to treat some non-equilibrium states
of quark ensemble (albeit with losing a covariance similar to the situation which
takes place in electrodynamics while one deals with electron-positron gas). But
here we are interested in the particular case of $\bar\mu=\mu$. Then the
corresponding derivative of specific energy $d w/d V$ might be presented as
\begin{equation}
\label{molodtsov:dwdV}
\frac{d w}{d V}=\int \frac{d{\bf p}}{(2\pi)^3}
\left(\frac{d n}{d\mu}\frac{d\mu}{dV}+\frac{d \bar n}{d\bar\mu}
\frac{d\bar\mu}{dV}\right)
\left[|p_4|\cos\theta-2G \sin \left(\theta-\theta_m\right)
\int\frac{d{\bf q}}{(2\pi)^3}
\sin\left(\theta'-\theta'_m\right)(n'+\bar n'-1)I\right].\nonumber
\end{equation}
Now expressing the trigonometric factors via dynamical quark mass
and exploiting Eq.(\ref{molodtsov:19}) we obtain the ensemble pressure as
$P=-\frac{E}{V}-V2N_c\int \frac{d{\bf p}}{(2\pi)^3}
\left(\frac{d n}{d\mu}\frac{d\mu}{dV}+\frac{d \bar n}{d\bar\mu}
\frac{d\bar\mu}{dV}\right)|P_4|$.
The requirement for mean charge conservation
$\frac{d \bar Q_4}{d V}=
\frac{\bar Q_4}{V}+V 2N_c\int \frac{d{\bf p}}{(2\pi)^3}
\left(\frac{d n}{d\mu}\frac{d\mu}{dV}-\frac{d \bar n}{d\bar\mu}
\frac{d\bar\mu}{dV}\right)=0$
provides us with an equation which interrelates the derivatives $d\mu/dV$,
$d\bar\mu/dV$. Apparently, the regularized expressions for mean charge of quarks
and anti-quarks are meant here. Dealing in a similar way with the requirement of
mean entropy conservation, $d \bar S/dV=0$, we receive another equation as
$\frac{\bar S}{2N_c~V^2}
=-\int \frac{d{\bf p}}{(2\pi)^3}\ln(n^{-1}-1)~\frac{d n}{d\mu}
\frac{d\mu}{d V} +\int \frac{d{\bf p}}{(2\pi)^3}
\ln (\bar n^{-1}-1)~\frac{d \bar n}{d\bar\mu}
\frac{d\bar\mu}{d V}$.
Substituting here $T\ln(n^{-1}-1)=-\mu+|P_4|$ and $T\ln (\bar n^{-1}- 1)=\bar\mu+|P_4|$
after simple calculations (keeping in mind that $\bar\mu=\mu$ and the charge
conservation) we have that $\int \frac{d{\bf p}}{(2\pi)^3}
\left(\frac{d n}{d\mu}\frac{d\mu}{dV}+\frac{d \bar n}{d\bar\mu}
\frac{d\bar\mu}{dV}\right)|P_4|=-\frac{\bar S T}{2N_c V^2}-\frac{\bar
Q_4\mu}{2N_c V^2}$. Finally it leads for the pressure to the following
expression $P=-\frac{E}{V}+\frac{\bar S~T }{V}+\frac{\bar Q_4~\mu}{V}$ (of course, the
thermodynamic potential is $\Omega=-P V$). At small temperatures the anti-quark
contribution is negligible and ~thermodynamic description can be grounded on
utilizing one chemical potential $\mu$ only. If the anti-quark contribution is getting
intrinsic the thermodynamic picture becomes complicated due to the presence of chemical
potential $\bar\mu$ with the condition $\bar\mu=\mu$ imposed. In particular, at
zero temperature the anti-quark contribution is absent and we might receive
$P= -{\cal E}+\mu~\rho_q$ where $\mu=[P_F^2+M^2_q(P_F)]^{1/2}$, $P_F$ is the
Fermi momentum and $\rho_q=N/V$ is the quark ensemble density.
\begin{figure}
\begin{minipage}{.45\textwidth}
\centering
\includegraphics[bb= 0 0 600 600, width=.9\textwidth]{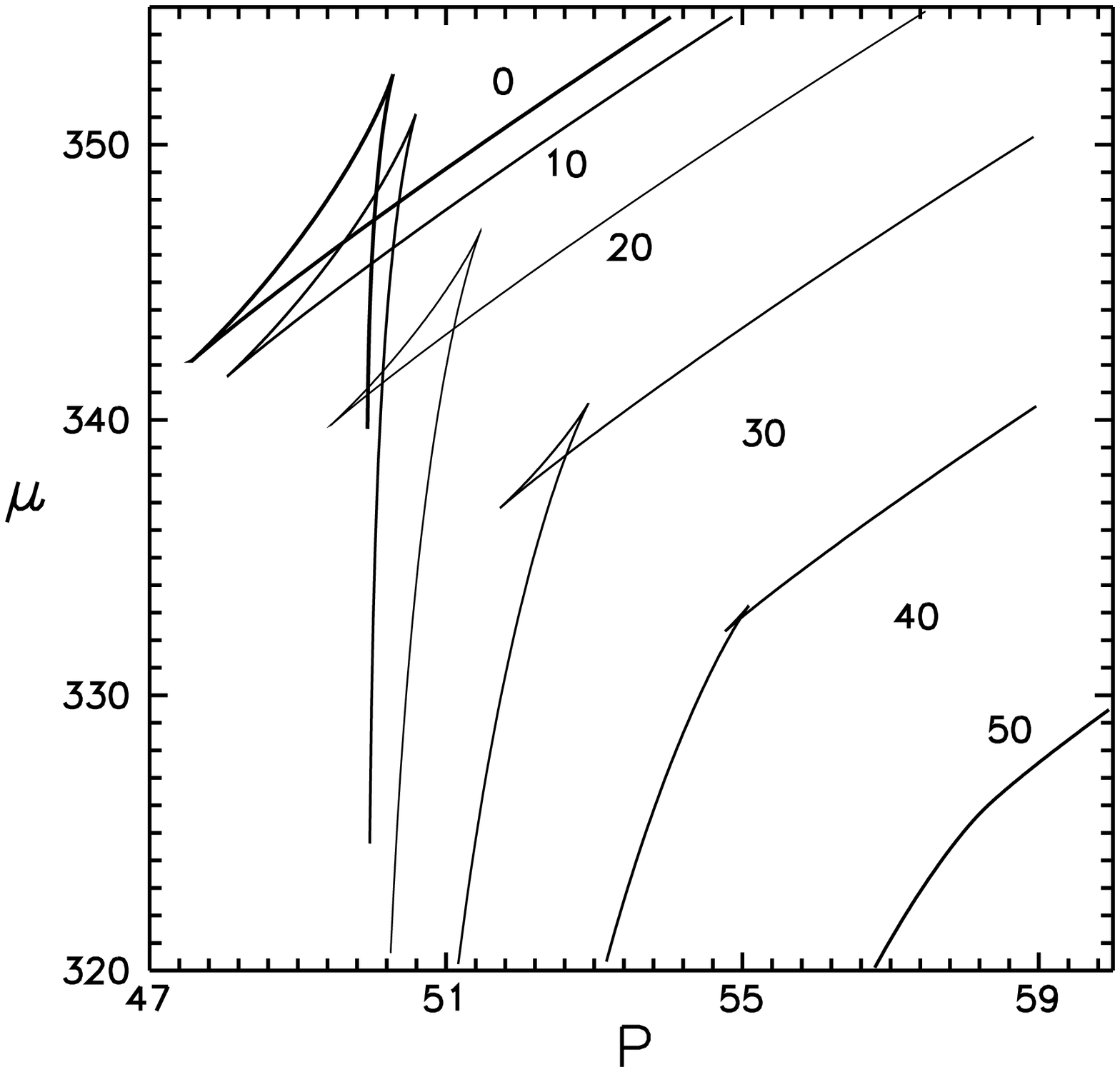}
\caption{The fragments of isotherms in Fig. \ref{molodtsov:f2}, see text.
Chemical potential $\mu$ (MeV) as a function of pressure $P$ (MeV/fm$^3$).
The top curve corresponds to the zero isotherm and following down with
spacing 10 MeV till the isotherm 50 MeV (the lowest curve).}
\label{molodtsov:f3}
\end{minipage}
 \rule{.05\textwidth}{0pt}
\begin{minipage}{.45\textwidth}
\centering
\includegraphics[bb= 0 0 600 600 ,width=.9\textwidth]{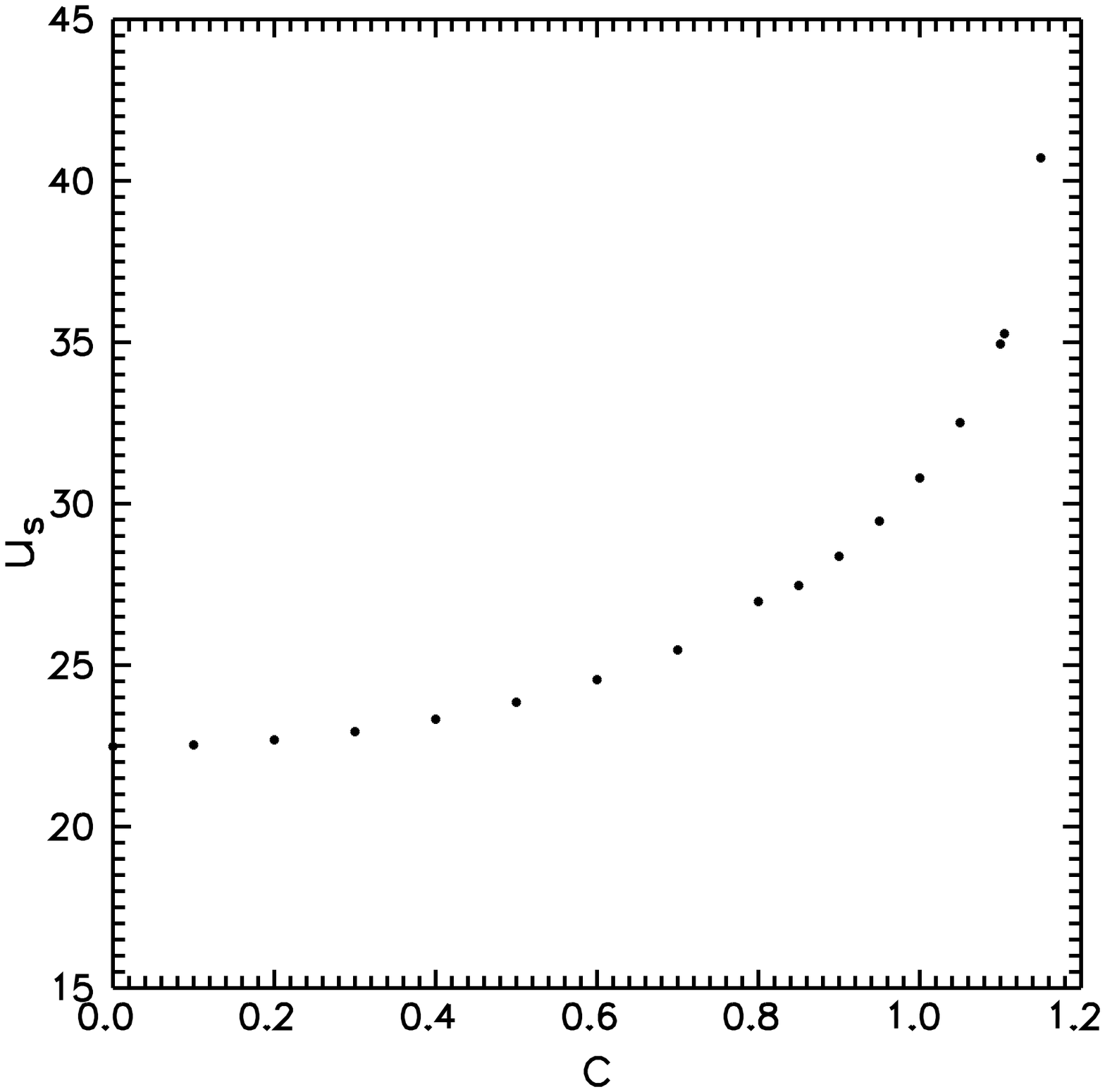}
\caption{The surface tension coefficient $u_s$ in MeV as a function
of parameter $c$ ($\zeta=c~\eta$) for the curve of stable kinks
(with $\eta\leq 1.2$).}
\label{molodtsov:f4}
\end{minipage}
\end{figure}

For clarity, we consider the NJL model  in this paper, i.e. the correlation
function behaves as the $\delta$-function in coordinate space. It is a well known fact
that in order to have an intelligent result in this model one needs to use a regularization
cutting of momentum integration in Eq. (\ref{molodtsov:17}). We adjust the standard set of
parameters \cite{molodtsov:5} here with $|{\bf p}|<\Lambda$, $\Lambda=631$ MeV, $m=5.5$ MeV
and $G\Lambda^2/(2\pi^2)=1.3$. This set of parameters at $n=0$, $\bar n=0$,
$T=0$ gives for the
dynamical quark mass $M_q=335$ MeV. Besides, it may be shown that the following
form of ensemble energy is valid at the extremals of functional (\ref{molodtsov:17})
$$E=E_{vac}+2 N_cV\int^\Lambda\frac{d{\bf p}}{(2\pi)^3}|P_4|
(n+\bar n)~,~~
E_{vac}=2 N_c V\!\!\! \int^\Lambda \frac{d{\bf p}}{(2\pi)^3}
(|p_4|- |P_4|)+2 N_c V \frac{M^2}{4 G}~.$$
It is easy to understand that this expression with the vacuum contribution
subtracted looks like an energy of a gas of relativistic particles and anti-particles with the
mass $M_q$ and coincides identically with that calculated in the mean field approximation.

Thus, we determine the density of quark $n$ and anti-quark $\bar n$ quasi-
particles at given parameters $\mu$ and $T$ from the second and third equations of system
(\ref{molodtsov:18}). From the first equation we receive the angle of quark and anti-quark pairing
$\theta$ as a function of dynamical quark mass $M_q$ which is handled as a parameter. The
evolution of chemical potential as a function of charge density ${\cal Q}_4=Q_4/(3V)$ (in the
units of charge/$fm^3$) with the temperature increasing is depicted in Fig.
\ref{molodtsov:f2} (factor 3 connects the quark and baryon matter densities). The top curve corresponds to
the zero temperature. The other curves following down have been calculated for the
temperatures $T=10$ MeV, ... , $T=50$ MeV with spacing $T=10$ MeV. As it was found in
Ref. \cite{molodtsov:ff} the chemical potential at zero temperature increases
first with the charge density increasing, reaches its maximal value, then decreases and at the
densities of order of normal nuclear matter density\footnote{At the Fermi momenta of
dynamical quark mass order.}, $\rho_q\sim 0.16/fm^3$, becomes almost equal to its vacuum value. Such
a behaviour of chemical potential results from the fast decrease of dynamical quark mass
with the Fermi momentum increasing. It is clear from Fig. \ref{molodtsov:f2} that the charge
density is still a multivalued function of chemical potential at the temperature slightly below
$50$ MeV. Fig. \ref{molodtsov:f2a} shows the ensemble pressure $P$ (MeV/fm$^3$) as the
function of charge density ${\cal Q}_4$ at several temperatures. The lowest curve corresponds to
the zero temperature. The other curves following up correspond to the temperatures
$T=10$ MeV, ... , $T=50$ MeV (the top curve) with spacing $T=10$ MeV. It is interesting to
remember now that in Ref. \cite{molodtsov:ff} the vacuum pressure estimate for the NJL model was
received as $40$---$50$ MeV/fm$^3$ which is entirely compatible with the results of the
conventional bag model. Besides, some hints at instability presence (rooted in the anomalous
behavior of pressure $dP/dn<0$) in an interval of Fermi momenta has been found. Fig.
\ref{molodtsov:f3} shows the fragments of isotherms of Fig. \ref{molodtsov:f2}, \ref{molodtsov:f2a} but in
the different coordinates (chemical potential --- ensemble pressure). The top curve is
calculated at the zero temperature, the other isotherms following down correspond to the
temperatures increasing with spacing 10 MeV. The lowest curve is calculated at the temperature 50 MeV.
The Fig. \ref{molodtsov:f3} obviously demonstrates a presence of the states on
isotherm which are thermodynamically equilibrated and have equal pressure and chemical
potential (see the characteristic Van der Waals triangle with the crossing curves). The calculated
equilibrium points are shown in Fig. \ref{molodtsov:f2a} by the dashed curve. The
intersection points of dashed curve with an isotherm are fixing the boundary of gas ---liquid phase
transition. The corresponding straight line $P=\mbox{const}$ which obeys the Maxwell rule
separates the non-equilibrium and unstable fragments of isotherm and describes a mixed phase
and appropriate critical temperature for the parameter we are using in this paper
turns out to be $T_{c}\sim 45.7$ MeV with the critical charge density as $\bar Q_4\sim
0.12$ charge/fm$^3$.
Usually the thermodynamic description is grounded on the mean energy functional
which is the homogeneous function of particle number like $E=N~f(S/N,V/N)$ (without vacuum
contribution). It is clear that such a description requires the corresponding subtractions to
be introduced, however, this operation does not change the final results considerably. It was
argued in Ref. \cite{molodtsov:ff} that the states filled up with quarks and separated
from the instability region look like 'natural construction material' to compose the
baryons and to understand the existing fact of equilibrium between vacuum and octet of stable
(in strong interaction) baryons\footnote{The chiral quark condensate for the filled up
state discussed develops the quantity about (100 MeV)$^3$ (at $T=0$), see \cite{molodtsov:ff},
that demonstrates the obvious tendency of restoring a chiral symmetry.}.

Apparently, our study of the quark ensemble thermodynamics produces quite
reasonable arguments to propound the hypothesis that the phase transition of chiral symmetry
(partial) restoration has already been realized as the mixed phase of physical vacuum and baryonic
matter{\footnote{Indirect confirmation of this hypothesis one could see, for example, in the existing
degeneracy of excited baryon states Ref. \cite{molodtsov:Gloz}.}}. However, it is clear our
quantitative estimates should not be taken as ones to be compared with, for example, the critical
temperature of nuclear matter which has been experimentally measured and equals to 15 -- 20 MeV.
Besides, the gas component (at $T=0$) has nonzero density (as $0.01$ of the normal nuclear
density) but in reality this branch should correspond to the physical vacuum, i.e. zero baryonic
density{\footnote{Similar uncertainty is present in the other predictions of chiral symmetry restoration
scenarios, for example, it stretches from 2 to 6 units of normal nuclear density.}}. In
principle, an idea of global equilibrium of gas and liquid phases makes it possible to formulate the
adequate boundary conditions at describing the transitional layer arising between the vacuum and
filled state and to calculate the surface tension effects.


The idea advanced would obtain substantial confirmation if it becomes possible
to claim an evidence of existing the transition layer at which the ensemble transformation
from one aggregate state to another takes place. As it was argued above the practical
parameter for describing an uniform phase (at a given temperature) is the mean charge
(density) of ensemble.
Thus, one can reconstruct all other characteristics, for example, a chiral
condensate, dynamical quark mass, etc. Analyzing the transition layer at zero temperature we assume
the parameters in the gas phase are approximately the same as at zero charge $\rho_g=0$, i.e. as
in the vacuum (ignoring the negligible distinctions in the pressure, chemical potential and
quark condensate). Then dynamical quark mass obtained has maximal value and for the parameter
choice of the NJL model it is $M=335$ MeV. From the Van der Waals diagram one may draw out that the
second (liquid) phase being in equilibrium with the gas phase develops the density $\rho_l=3\times
0.185$ ch/fm$^3$. The detached factor 3 here relates the magnitudes of quark and baryon densities. The
quark mass in this phase is approximately $\stackrel{*}{M}\approx 70$ MeV (we are dealing further
with the simple one-dimensional picture).

Usually an adequate description of heterogeneous states can be developed basing
on the mean field approximation \cite{molodtsov:Lar}, specifically for our case, by dealing with
the corresponding effective quark-meson Lagrangian (a sort of the Ginzburg-Landau functional)
\begin{equation}
\label{molodtsov:mesons}
{\cal L}=-\bar q~(\hat \partial +M)~q- \frac12~(\partial_\mu \sigma)^2
-U(\sigma)-\frac14~F_{\mu\nu}F_{\mu\nu}-\frac{m_v^2}{2}~V_\mu V_\mu-
g_\sigma~\bar q q~\sigma+i g_v~\bar q~\gamma_\mu~q~ V_\mu~,\nonumber
\end{equation}
where $F_{\mu\nu}=\partial_{\mu} V_\nu-\partial_{\nu} V_\mu$,
$U(\sigma)=\frac{m_\sigma^2}{2} \sigma^2+ \frac{b}{3}\sigma^3
+\frac{c}{4}\sigma^4$, $\sigma$ is the scalar field, $V_\mu$ is the field of
vector mesons, $m_\sigma$, $m_v$ are the masses of scalar and vector mesons and $g_\sigma$,
$g_v$ are the coupling constants of quark-meson interaction. The $U(\sigma)$ potential includes the
nonlinear terms of sigma-field interactions up to the fourth order, for example. For the sake of
simplicity we do not include the contribution coming from the pseudoscalar and axial-vector mesons.

We are not going beyond well elaborated (and quite reliable) one loop
approximation (\ref{molodtsov:mesons}), although recently the considerable progress was
reached in scrutinizing the non uniform quark condensates by utilizing the powerful methods of exact
integration \cite{molodtsov:KN}. We believe it is more practical to adjust
phenomenologically the parameters of effective Lagrangian being guided also by transparent physical picture. It is
easy to see that handling one loop approximation  actually we have the Walecka model
\cite{molodtsov:wal} but applied for the quarks. In what follows we are working with the notations of that model
hoping it does not lead to the misunderstandings. In the context of our deliberation Eq.
(\ref{molodtsov:mesons}) can be interpreted in the following way. Each phase, in some extent, might be
considered as an excited state as to its relation with another phase which requires an additional (besides a
charge density) set of parameters just as the meson fields for describing and those fields characterize
the measure of deviation from the equilibrium state. Then the key question becomes whether it
is possible to adjust the effective Lagrangian parameters  of (\ref{molodtsov:mesons}) in order to
obtain the solutions in which the quark field interpolates between the quasi-particles in the gas
(vacuum) phase and in the quasi-particles of the filled up state. The density of ensemble of the filled up
states should asymptotically approach an equilibrium value of $\rho_l$ and turn to zero value
in the gas phase (vacuum).

Taking the parameterization of the potential $U(\sigma)$ as
$b_\sigma=1.5~m_\sigma^2~(g_\sigma/M)$,
$c_\sigma=0.5~m_\sigma^2~(g_\sigma/M)^2$ we come to the sigma model and the
choice $b=0$, $c=0$
results in the Walecka model. As to the application for nuclear matter the
parameters $b$ and $c$  demonstrate essentially the model dependent character and are different from the
parameter values of the sigma model. They are phenomenologically adjusted with requiring an accurate
description of the saturation property. On the contrary, for the quark Lagrangian (\ref{molodtsov:mesons}) we
could intuitively anticipate some resemblance with the sigma model and, hence, introduce two
dimensionless parameters $\eta$, $\zeta$ as $b=\eta~b_\sigma$, $c=\zeta^2~c_\sigma$ which characterize some
fluctuations of the effective potential. Then the scalar field potential is presented by the following form
$U(\sigma)=\frac{m_\sigma^2}{8}\frac{g_\sigma^2}{M^2}
\left(4\frac{M^2}{g_\sigma^2}+4\frac{M}{g_\sigma}
\eta\sigma+\zeta^2\sigma^2\right)\sigma^2$.
The meson and quark fields are defined by the following system of the stationary equations
\begin{eqnarray}
\label{molodtsov:sys}
&&\Delta~ \sigma -
m_\sigma^2~\sigma=b~\sigma^2+c~\sigma^3+g_\sigma~\rho_s~,\nonumber\\
&&\Delta ~V - m_v^2~ V=-g_v~\rho~,\\
&&( {\bf \hat\nabla}+\stackrel{*}{M})~q=(E-g_v~V)~q~\nonumber
\end{eqnarray}
where $\stackrel{*}{M}=M+g_\sigma\sigma$ is the running value of dynamical quark
mass, $E$ stands for the quark energy and $V=-iV_4$. The density matrix describing the quark ensemble at
$T=0$ has the form $ \xi (x)=\int^{P_F}\frac{{d\bf p}}{(2\pi)^3}q_{\bf p}(x)\bar q_{\bf
p}(x)$ where ${\bf p}$ is the quasi-particle momentum and the Fermi momentum $P_F$ is defined by the ensemble
chemical potential. The densities $\rho_s$, $\rho$ in the right hand sides of equations
(\ref{molodtsov:sys}) equal (by definition)
to $\rho_s(x)=Tr\left\{\xi(x),1 \right\}$,
$\rho(x)=Tr\left\{\xi(x),\gamma_4\right\}$. Here we confine
ourselves to the Thomas--Fermi approximation while describing the quark ensemble.
Then the densities in which we are interested in are given with some local Fermi momentum $P_F(x)$ as
$\rho= \gamma\int^{P_F}\frac{d {\bf p}}{(2\pi)^3}=\frac{\gamma}{6\pi^2}P_F^3$
$\rho_s=\gamma\int^{P_F}\frac{d {\bf p}}{(2\pi)^3}
\frac{\stackrel{*}{M}}{E}$ where $\gamma$ is the quark gamma factor ($\gamma=2
N_c N_f$, $N_c$ is the number of colours, $N_f$ number of flavours), $E=({\bf p}^2+\stackrel{*}{M}^2)^{1/2}$.
By definition the ensemble chemical potential does not change and it leads to the situation in which the
local value of Fermi momentum is defined by the running value of dynamical quark mass and vector field as
$\mu=M=g_v~V+(P_F^2+\stackrel{*}{M}^2)^{1/2}$. The details of tuning the Lagrangian parameters (\ref{molodtsov:mesons}) can be
found in \cite{molodtsov:arxiv}. The point of our attraction here is the surface tension coefficient
\cite{molodtsov:Bog} $u_s=4\pi~r_o^2~\int_{-\infty}^{\infty}dx~
\left[ {\cal E}(x)-\frac{{\cal E}_l}{\rho_l}~\rho(x)\right]$,
here ${\cal E}_l$ is the energy density in the liquid phase. The parameter
$r_o$ is discussed below. In the Thomas--Fermi approximation ${\cal E}(x)=\gamma\int^{P_F(x)}\frac{d {\bf
p}}{(2\pi)^3} [{\bf p}^2+\stackrel{*}{M}(x)]^{1/2}+\frac12 g_v \rho(x)V(x)-\frac12
g_\sigma\rho_s(x)\sigma(x)$. The surface tension coefficient $u_s$ in MeV for the curve of stable kinks (see
the details in Ref. \cite{molodtsov:arxiv}) with parameter $\eta\leq 1.2$ as the function of
another parameter $c$ ($\zeta=c~\eta$) is depicted in Fig. \ref{molodtsov:f4}.


Above results lead us to put the challenging question about the properties of
finite quark  systems or droplets of quark liquid which are in equilibrium with the vacuum
state. As a droplet here we imply the spherically-symmetric solution of the equation system
(\ref{molodtsov:sys}) for $\sigma(r)$ and $V(r)$ with the obvious boundary conditions
$\sigma'(0)=0$ and $V'(0)=0$ in the origin (the primed variables denote the first derivatives over $r$) and rapidly
decreasing at the large distances $\sigma \to 0$, $V \to 0$ when $r\to \infty$.
\begin{figure}
\begin{minipage}{.45\textwidth}
\centering
\includegraphics[bb= 0 0 600 600, width=.9\textwidth]{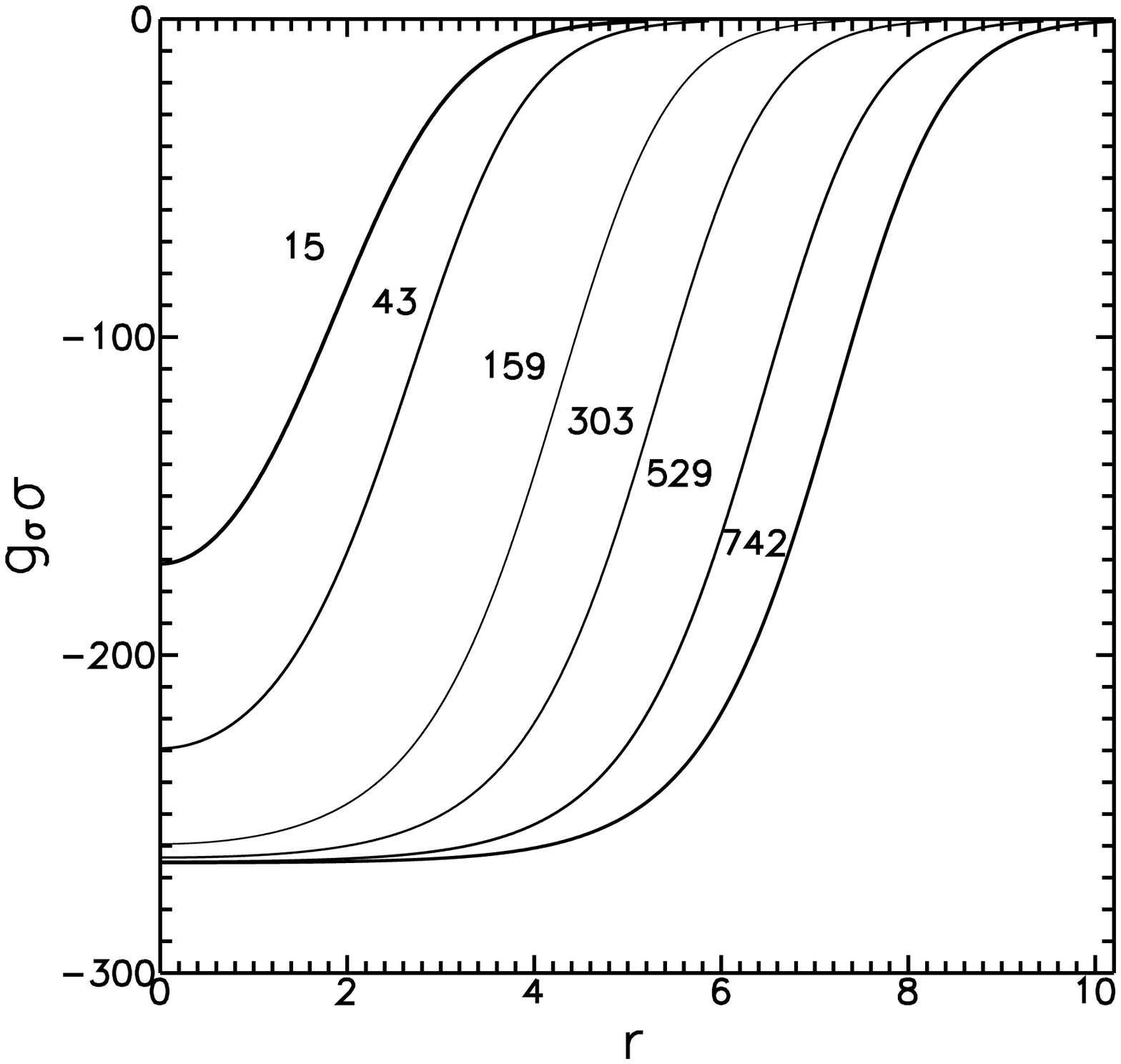}
\caption{$\sigma$-field (MeV) as a function of the distance $r$ (fm)
for several solutions of the equation system (\ref{molodtsov:sys}) which are
characterized by the net quark number $N_q$ written to the left of each curve.
}
\label{molodtsov:f7}
\end{minipage}
 \rule{.05\textwidth}{0pt}
\begin{minipage}{.45\textwidth}
\centering
\includegraphics[bb= 0 0 600 600 ,width=.9\textwidth]{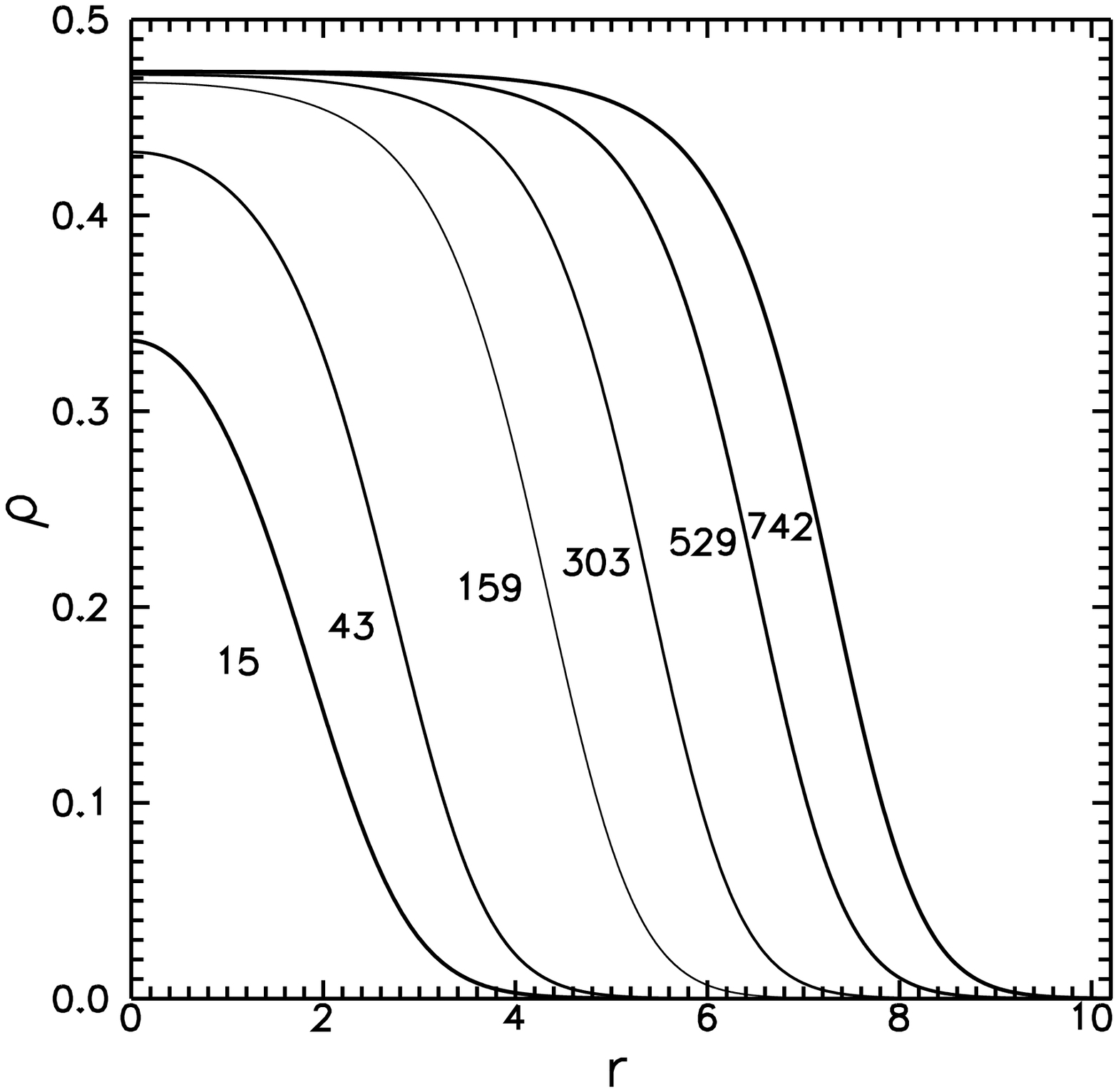}
\caption{Distribution of the quark density $\rho$ (ch/fm$^3$)
for the corresponding solutions presented in the Fig. \ref{molodtsov:f7}.
}
\label{molodtsov:f7a}
\end{minipage}
\end{figure}
Fig. \ref{molodtsov:f7} shows the set of solutions ($\sigma$-field (MeV)) of the
equation system (\ref{molodtsov:sys}) at number of flavors $N_f=1$. Fig. \ref{molodtsov:f7a}
presents the corresponding distributions of ensemble density  $\rho$ (ch/fm$^3$). The Table 1 exhibits the
results of fitting the density $\rho(r)$ with the Fermi distribution $\rho_F(r)=\frac{\widetilde
\rho_0}{1+e^{(R_0-r)/b}}$ where $\widetilde \rho_0$ is the density at the origin, $R_0$ is the mean size
of the droplet and the parameter $b$ determines the thickness of surface layer $t=4\ln(3) b$. Besides,
the coefficient $r_0$ which is included in the definition of surface tension coefficient, $R_0=r_0
N_q^{1/3}$ is also presented together with the characteristic values of the $\sigma$-meson mass and
the coefficient $\eta$ at which all this values were obtained.
\begin{center}
{\underline{Table 1}}. Results of fitting by the Fermi distribution
($N_f=1$).
\\\vspace{0.3cm}
\begin{tabular}{|l|c|c|c|c|c|c|l|}
\hline
$N_q$ &$\widetilde
        \rho_0$
       (ch/fm$^3$) &$R_0$
                   (fm)   &$b$
                          (fm$^{-1})$&$t$ (fm) &$r_0$ (fm)&$m_\sigma$
(MeV)&
$\eta$ \\\hline
$15$  &$0.34$     &$1.84$ &$0.51$    &$2.24$   &$0.74$    &$351$
&
$0.65$    \\
$43$  &$0.43$     &$2.19$ &$0.52$    &$2.28$   &$0.75$    &$384$
&
$0.73$    \\
$159$ &$0.46$     &$4.19$ &$0.52$    &$2.29$   &$0.77$    &$409$
&
$0.78$    \\
$303$ &$0.47$     &$5.23$ &$0.52$    &$2.29$   &$0.78$    &$417$
&
$0.795$    \\
$529$ &$0.47$     &$6.37$ &$0.52$    &$2.27$   &$0.79$    &$423$
&
$0.805$    \\
$742$ &$0.47$     &$7.15$ &$0.52$    &$2.27$   &$0.79$    &$426$
&
$0.81$     \\
\hline
\end{tabular}
\end{center}
The curves plotted in the Fig. \ref{molodtsov:f7} and results of Table 1 allows
us to conclude that the density distributions at  $N_q\ge 50$ correspond quite well to
the data typical for the nuclear medium. The thicknesses of transition layers
are also similar. The coefficient $r_0$ with the factor $3^{1/3}$ included is in full
correspondence with nuclear one. The values of
the $\sigma$-meson mass turn out to be quite reasonable as well.
Although at small quark numbers in the droplet the
corresponding behaviors become essentially different. We know experimentally that in the nuclear matter
one can observe some increase of the ensemble density which is quite considerable for the Helium
and is much higher than the normal nuclear matter density for the Hydrogen.  One may criticize us in this point
because working within the Thomas--Fermi approximations becomes hardly justified at the small
number of quarks and it is necessary to handle the solution of equation system
(\ref{molodtsov:sys}). However, fortunately, the exploration we are interested in has been performed in the
chiral soliton model of nucleon \cite{molodtsov:BB}. It has been demonstrated there that adding the
contributions of pseudo-scalar and axial-vector fields to the Lagrangian (\ref{molodtsov:mesons})
leads to reasonably good description of nucleon and $\Delta$. The interesting remark here is that
the soliton solutions obtained in \cite{molodtsov:BB} could be interpreted as a 'confluence' of two
kinks. Each of those kinks develops the restoration of chiral symmetry in a sense that the scalar
field is approaching its zero value at the distance $\sim 0.5$ fm from the kink center. Actually, one
branch corresponds to the solution with the positive value of the dynamical quark mass and another
branch presents the solution with negative dynamical quark mass (in three-dimensional picture the
pseudo-scalar fields appears just as a phase of chiral rotation from
positive to negative value of quark mass).

\acknowledge{Acknowledgements}{We are grateful to the meeting organizers and
personally Professor V. Skalozub for a hospitality and excellent conditions for efficient
work.
}

\end{article}
\label{pgs1}
\end{document}